\documentclass[11pt]{article}
\usepackage{graphics}
\usepackage{feynmf}
\title{DUAL INTERPRETATIONS OF PION CLOUDS AT RHIC}
\author{JACK L. URETSKY \\ High Energy Physics Division,
Argonne National Laboratories}
\date{\today}
\begin{document}\setlength{\unitlength}{.7mm}
\hfill ANL-HEP-PR-06-62
\vspace{7 mm}
\begin{center}
Dual Interpretations of Pion Clouds at RHIC

\vspace{3 mm}
             JACK L. URETSKY,
 
 High Energy Physics Division,Argonne National Laboratories
\end{center}
\section{ABSTRACT}
A gauge theory of pions interacting with rho-mesons at elevated temperatures is used to calculate the pressure in a hot pion gas. No reference is made to 
the pion's status as a QCD Goldstone boson. The role of the pion is merely 
that of a carrier of an SU(2) symmetry, gauged to create a vector-meson 
interaction, the rho playing the role of the interacting vector particle.  
The results are in rough agreement with much more elaborate calculations, 
both of the purely hadronic variety, and those that invoke quark-gluon degrees 
of freedom. The quark-gluon and purely hadronic calculations seemingly lead 
to very similar predictions which are in accord with receent data from RHIC.  
The results motivate the question as to whether the two descriptions are 
dual to each other in the sense of being alternate models, each sufficient 
to explain the observed data.
\section{INTRODUCTION}
Experimental groups at RHIC have recently summarized their results from 
three years of measurements on gold-gold collisions with center-of-mass 
energies per nucleon pair of 130 and 200 GeV.  I shall focus, 
for definiteness, on results from the 200 GeV per nucleon pair runs 
reported by the PHENIX experiment~\cite{ph:sum}.

Heavy ion collisions, even before the RHIC experiments, were found to 
produce clouds of hundreds of particles, with a predominance of pions, 
per collision \cite{NA 49}.  One can imagine that, in accordance with 
Bjorken's description \cite{BJ:collide}, the colliding ions pass through 
each other like porous pancakes and vanish down the beampipe, leaving 
behind a fireball of pions and other hadrons.  Like-pion correlation 
measurements at RHIC reported in reference \cite{ph:sum} (Section 3.4) 
suggest that the fireball has about the size of a 
gold nucleus (radius $\approx 6$ fm.  There are about 
1000 pions per unit rapidity produced in a central collision of gold nuclei 
with $\sqrt{s_{NN}}=200$ GeV at RHIC \cite{ph:yld}, leading to an 
estimated pion density in the fireball of about $2$ $ fm^{-3}$.
These fireballs are a new form of matter (characterized by Shuryak 
\cite{pi:liq} as ``the pion liquid'' - partially in recognition of the high 
density of particles in the clouds).

A task for the theorist, faced with the ephemeral presence of this rare 
form of matter, is to obtain its equation of state and investigate its 
observable properties.  A number of authors have attacked various aspects 
of this task during the past several decades, see {\em e.g.} \cite{pi:liq}, \cite{Baym:83}, \cite{Gerber:90}, \cite{Welke:90}, 
\cite{Rapp:95}, \cite{Dobado:98}, and papers cited therein.  The common 
approach is through the relativistic thermodynamics of an interacting pion 
gas. Reference \cite{Rapp:95} is based upon a hadronic effective Lagrangian 
for the pionic interactions, with in-medium modification of the single pion 
propagator. Reference \cite{Dobado:98} bases its $\pi \pi $ interaction on chiral perturbation theory.  Both approaches reproduce experimental (zero-temperature) scattering data so that thermodynamic variables calculated in the two approaches are expected to, and appear to, agree by virtue of a theorem of Dashen, Ma and Bernstein \cite{Dashen:1969}. 

Reference \cite{Dobado:98} includes density considerations by noting 
that pi-pi scattering is almost purely elastic in the energy range of interest, 
so that pion number can be treated as a conserved quantity.  This observation is used to justify the inclusion of a chemical potential corresponding to pion number.  See also, Welke, {\em et al.} \cite{Welke:90}

The approach in the present paper exploits a remark made long ago in reference \cite{Gasser:1984}. The authors say of their chiral perturbation theory results for pi-pi scattering (p. 192): ``We show that in a systematic low-energy expansion in powers of the momenta the $\rho $ does not play any special role - its presence only manifests itself in the values of the low energy constants''.  But we learn from reference \cite{Dashen:1969} that the knowledge of the S-matrix is sufficient for calculating the grand canonical potential of any gaseous system. I take this as meaning that a thermal field theory of pi's and rho's should provide a ``pretty good'' description of a hot pion cloud.  The meaning of ``pretty good'' may be gleaned from the fact that $\rho $-exchange in the $\pi \pi $- scattering t-channel gives respective I-spin $0$ and $2$ scattering lengths of $.36 m_{\pi }^{-1}$ and $-.18 m_{\pi }^{-1}$, whereas a recent precision determination of the lifetime of the pionium atom by the Dirac Collaboration \cite{Adeva:05} gave 
$|a_{0} - a_{2}|= .264^{+.033}_{-.020}\: m_{\pi }^{-1}$.  

\section{The Lagrangian}
As proposed in 1967 by Lee and Nieh \cite{Lee:67} and Weinberg \cite{SW:67}, I start by providing the pion field with a local SU(2) gauge transformation to generate a gauge-invariant coupling to an isovector vector field.  I then break the SU(2) symmetry by providing the vector field with the mass $M$ of the $\rho $ - meson.  The resultant broken symmetry is sometimes referred to as a ``hidden local gauge symmetry'' \cite{Bando:85}.  The ``hidden symmetry'' manifests itself by preserving coupling-constant relations, despite the breaking.

Introducing the vector field $V_{\mu}$ and the covariant derivative 
\begin{equation}
\delta_{\mu} = d_{\mu} - ig\vec{V_{\mu}}\cdot \vec{T}\\
\label {eq:cov}
\end{equation}
where
\begin{equation}
\vec{T} \times \vec{T} = i\vec{T}
\end{equation}
the pion Lagrangian density is 
\begin{equation}
\mathcal{L}_{\pi }= -\frac{1}{2}( \delta \vec{\pi} \cdot \delta\vec{\pi} -m^{2} \vec{\pi} \cdot \vec{\pi})\\
\label{eq:pi}
\end{equation}
where the vector symbols refer to SU(2) space.  The total Lagrangian density is then
\begin{equation}
\mathcal{L}_{TOTAL} = \mathcal{L}_{\pi} - \frac{1}{4}\vec{F}_{\mu \nu} \cdot \vec{F}^{\mu \nu} - \frac{1}{2} M^{2}(\vec{V}_{\mu} \cdot \vec{V}^{\mu})
\label{eq:tot}
\end{equation}
with
\begin{equation}
\vec{F}_{\mu \nu} \equiv d_{\mu}\vec{V}_{\nu} - d_{\nu}\vec{V}_{\mu} + g\vec{V}_{\mu}\times \vec{V}_{\nu}
\label{eq:Fmunu}
\end{equation}
The interaction Lagrangian density is the part of $\mathcal{L}_{TOTAL}$ that is proportional to $g$ and $g^{2}$.\\
\begin{equation}
\mathcal{L}_{INT} = -g\vec{V}^{\mu }\cdot [\vec{\pi }\times  d_{\mu }\vec{\pi }] - \frac{1}{2}g^{2}\vec{V}^{\mu }\cdot [\vec{\pi }
\times (\vec{V}_{\mu }\times \vec{\pi })]
\label{eq:int}
\end{equation}

\section{THERMAL MASS AND WIDTH OF THE RHO}
The order $g^{2}$ contribution from the interaction described by Eq. \ref{eq:int} to the thermal $\rho $-polarization tensor can be calculated using the methods described by Kapusta \cite{Kap:89}, Le Bellac \cite{le:96}, and Pisarski \cite{Pis:87}  The relevant Feynman diagrams are shown in Fig. 1.  I evaluate them in the Matsubara formalism \cite{Mat:55}.
\begin{center}
  \begin{fmffile}{samplepics}
\begin{fmfgraph*}(75,50)
\fmfpen{thick}
\fmfleft{i} \fmfright{o}
\fmf{plain}{i,v1} \fmf{plain}{v2,o}
\fmf{dots,left}{v1,v2,v1}
\fmflabel{Fig. 1}{v2}
\end{fmfgraph*}
\begin{fmfgraph*}(75,50)
\fmfpen{thick}
\fmfleft{i} \fmfright{o}
\fmf{plain}{i,v1,o}
\fmf{dots,right}{v1,v1}
\end{fmfgraph*}
\end{fmffile} \\
\begin{figure}
\protect\caption{Feynman diagrams for $\rho $-polarization tensor; $\rho $- full line, pion- dotted line \protect}
\end{figure}
Fig.1 Feynman diagrams for the second order $\rho $-polarization tensor.
\end{center}

The $\rho$-propagator in Euclidean momentum space is:
\begin{equation}
D(q)_{\mu \nu } = L(q)_{\mu \nu }\frac{1}{q^{2}+M^{2}}
\label{eq:prop}
\end{equation}
with
\begin{equation}
L(q)_{\mu \nu } = \delta_{\mu \nu } - \frac{q_{\mu }q_{\nu }}{q^{2}}
\label{eq:landau}
\end{equation}
The choice of Landau gauge for the propagator insures that the $\rho $ will have three components both on- and off-shell.  

The squared mass of the $\rho$ at temperature $T$, including one-loop order pion contributions, is given by:
\begin{eqnarray}
\mathcal{M}(p)^{2} -M^{2}& =  &\frac{-g^{2}}{3(2\pi )^{3}}T^{2}\sum_{n,n^{^{\prime }}} \langle \int d^{3}kd^{3}k^{\prime } \Delta (k)\Delta (k^{\prime })\int_{0}^{\beta }dy [e^{iy(p_{0}-k_{0}-k^{\prime  } _{0})}\delta (\vec{p}-\vec{k}-\vec{k^{\prime  }} ) \nonumber \\
                         &    & \frac{p^{2}(k-k^{\prime})^{2}]-[q\cdot (k-k^{\prime })]^{2})}{p^{2}}+[2D^{\nu }_{\nu }(0)+\frac{3}{2}\Delta(0)+\frac{p_{\mu }p_{\nu }}{3p^{2}}D^{\mu \nu }(0) \rangle \nonumber \\
\label{eq:mass1}
\end{eqnarray}
where the argument ``$0$'' in a propagator refers to the spacetime origin, $\Delta(k)$ is (inconsistently) the momentum-space pion propagator, and $D^{\mu \nu }(p)$ is the momentum-space $\rho $-propagator.  The indices $n, n^{\prime  } $ refer respectively to the discrete Matsubara \cite{Mat:55} modes of the fourth components of the Euclidean four-vectors ``$k. k^{\prime  } $''

Further details of the $\rho $-mass calculation are provided in Appendix A.  The temperature dependence of the mass and width of $\rho$ -mesons at rest in a heat bath are repectively shown in Fig. 2.

\begin{center}
\begin{figure}
\includegraphics*[0in,0in][6.5in,4.5in]{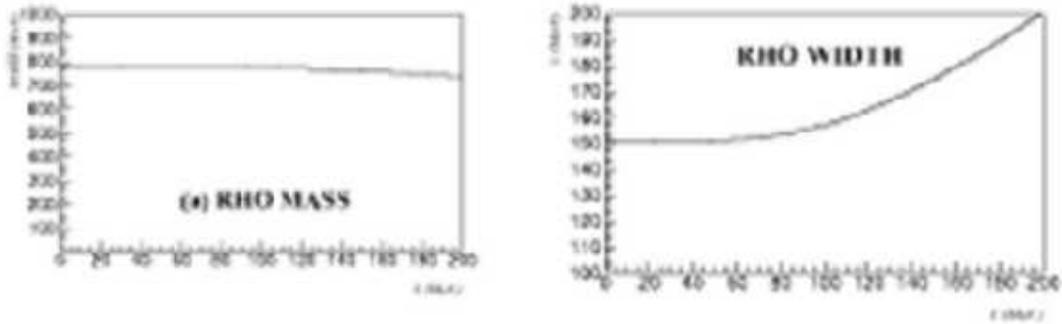}
\protect\caption{Mass and width of the $\rho $ vs temperature \protect}
\end{figure}
\end{center}
 It is evident from Eq. \ref{eq:mass} that the effect of the heat bath is a negative contribution to the mass of the $\rho $ which turns out to be small at temperatures below $200\: MeV$.  The width is Bose-enhanced, as previously discussed in \cite{Urban}, in accordance with Eq. \ref{eq:width}.  These results are in at least qualitative agreement with results from the NA60 experiment at the CERN SPS \cite{NA60}

\section{EQUATION OF STATE OF A HOT PION GAS}
The pressure $p$ of a hot pion gas, to second order in the $\pi - \rho $ coupling, may be written in the form (\cite{Kap:89}, \cite{le:96})
\begin{equation}
\frac{p}{T^{4}} = \Lambda _{0} + 3g^{2}\Lambda
\label{eq:press}
\end{equation}
The free-pion contribution $\Lambda _{0}$ is given by
\begin{equation}
\Lambda _{0} = \frac{\beta ^{4}}{2\pi ^{2}}\int_{0}^{\infty }n(\beta \omega )\frac{k^{4}dk}{\omega }
\label{eq:freepi}
\end{equation}
with $\omega $ denoting the energy of a pion of momentum $k$, $\beta $ the inverse of the temperature $T$, and $n(x)$ the Bose-Einstein distribution function $(e^{x}-1)^{-1}$.

The order $g^{2}$ pion-contribution to $\lambda $ in Eq. \ref{eq:press} can be divided into two contributions, as shown in Fig. 1.  The corresponding expression may be written
\begin{equation}
\Lambda = \Sigma \int \frac{d^{3}kd^{3}p}{(2\pi )^{6}}[A(p,k)+B(p,k)]
\label{eq:lam}
\end{equation}
with 
\begin{eqnarray}
A(p,k) & = &\frac{T^{3}}{2}\int d^{3}k^{\prime }\delta (\vec{k}+\vec{k^{\prime}}+\vec{p})\int_{0}^{\beta }d\tau e^{i(p_o)+k_{0}+k^{\prime }_{0})\tau }[ \nonumber \\
       &   & \Delta (k)\Delta (k^{\prime })D^{\mu \nu}(p)(k-k^{\prime })_{\mu }(k-k^{\prime })_{\nu }] \nonumber \\
\label{eq:A}
\end{eqnarray}
and
\begin{equation}
B(p,k) = T^{2}D^{\nu }_{\nu }(0)\Delta (0)
\label{eq:B}
\end{equation}
The summation is over the respective Matsubara modes of the fourth components of $p,k,k^{\prime }$,  and I have symmetrized in the momenta $k$ and $k^{\prime}$. Some details of the calculation are in Appendix B. A graph of the resulting pressure-temperature relation is shown in Fig. 3, where it is compared with the corresponding results obtained by Rapp and Wambach. \cite{Rapp:95}.   Fig. 3 shows that the simple model presented here overestimates the pressure by about $30\%$ at a temperature of $170 \:MeV$ compared with the model in reference \cite{Rapp:95}
\begin{center}
\begin{figure}
\includegraphics[0in,0in][4in,3in]{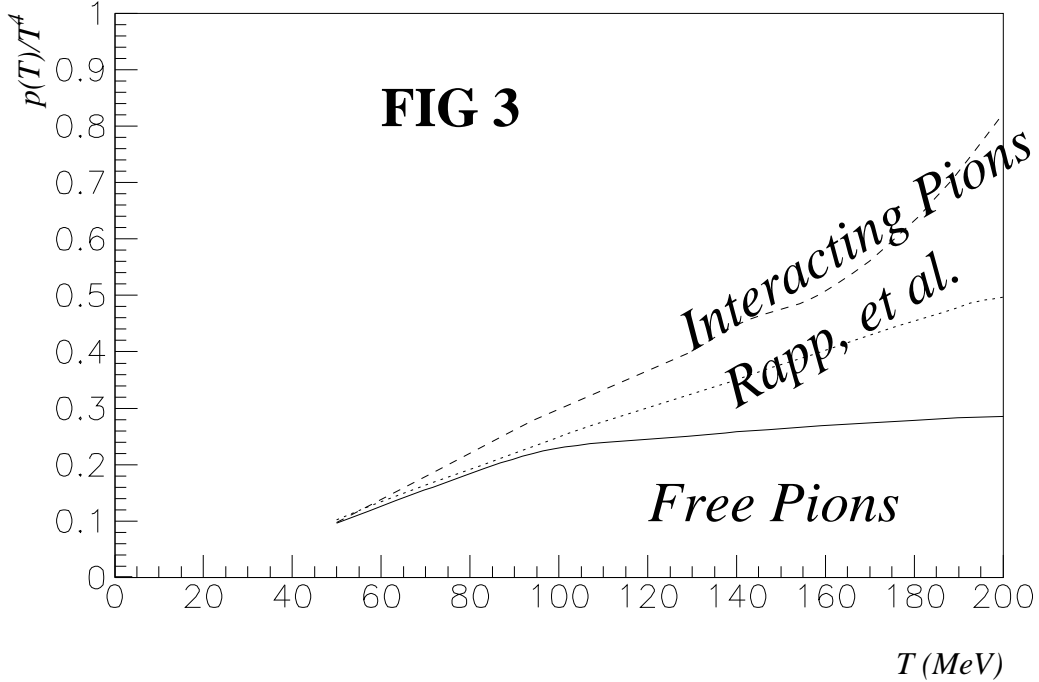} 
\protect\caption{Pressure v Temperature in a pion gas; dashed line-this calculation, dotted line- Rapp \& Wambach \protect}
\end{figure}
\end{center}

The result obtained in ref. \cite{Rapp:95} is almost identical with a QCD-related calculation by Nikonov, {\em et al.} \cite{Nikonov:98}.  That calculation treats the ``fireball'' aftermath of the heavy-ion collision as a ``mixed-phase'' combination consisting of both a quark-gluon phase and a hadronic phase.
  
\section{The Pion Gas at RHIC}
The aftermath of a heavy ion collision is an expanding fireball of pions and other debris.  Transverse-momentum spectra of $\pi 's,\ K's\ p's\ and\ \bar{p}'s$ are plotted and tabulated in ref. \cite{ph:yld} for $\sqrt{s_{NN}}=200$ GeV gold on gold collisions at RHIC. The transverse momentum of a debris particle is compounded from the expansion of the cloud and the thermal motion of the particles in the cloud.  Siemens and Rasmussen \cite{Siemens 79} pointed out three decades ago that the heavy particles in the expanding cloud can be used to measure the expansion velocity of the cloud.  Their point was that the peak of a heavy-particle spectrum represents the momenta of those particles with zero transverse thermal velocity.

Fig. 4 is a plot of the transverse-momentum spectrum of $K^{+}$ mesons taken from Table XVII of ref. \cite{ph:yld} (centrality 0-5\%, error bars omitted),  
\begin{center}
\begin{figure}
\includegraphics[0in,0in][3in,2in]{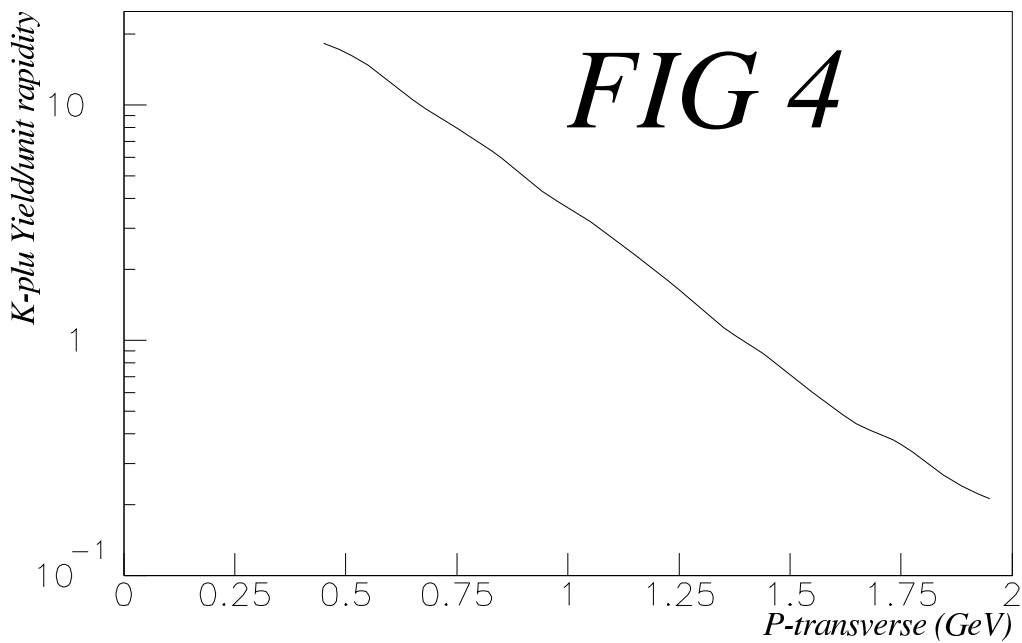}
\protect\caption{Transverse momentum spectrum of $K^{+}$ mesons from Au+Au collisions at $\sqrt{s_{NN}}$ GeV at RHIC \protect}
\end{figure}
\end{center}
  Although the peak is not included in the reported measurement, we can roughly estimate that it would be near, or a bit below, .35 GeV.  Taking the charged-K mass as .5 GeV, results in a $\beta ^{2}$ of $.32$, or a bit less, for the squared speed (divided by c) of the expanding cloud.  This result is close to the limit of $1/3$ for a gas of highly relativistic particles \cite{Wein:72}, and about equal to the expansion velocity of a free-pion gas at a temperature in excess of about $150$ MeV, as reported by Welke, {\em et al.} \cite{Welke:90}.  

The observed slope of the kaon spectrum corresponds to an apparent temperature 
of $T_{app} \approx 290$ MeV.  Correcting this value by transforming to the frame of the expanding gas cloud result \cite{Siemens 79} results in a Kaon temperature $T=\sqrt{\frac{1-\beta }{1+\beta}}T_{app}\approx 150$ MeV, near the range of temperatures estimated in ref. \cite{ph:sum}.
\section{CONCLUSIONS}
The implicit questions underlying the present work are: What are the limits of the theorem of Dashen, {\em et al.} \cite{Dashen:1969}, that makes the equation of state of a collecttion of hadrons an offspring of a hadronic S-matrix? Can purely hadronic descriptions of the thermodynamics of the ``fireball'' created by heavy ion collisions account, at least qualitatively, for the results observed at RHIC? \cite{ph:sum}? Or, must quark-gluon degrees of freedom be invoked to properly describe the RHIC results?

The questions are posed in the context of a simple model, a gauge theory of pions interacting with $\rho$-mesons at elevated temperatures. No reference is made to the pion's status as a QCD Goldstone boson, in contradistinction to the works cited in the Introduction. The role of the pion is merely that of a carrier of an SU(2) symmetry, gauged to create a vector-meson interaction, the $\rho $ playing the role of the interacting vector particle.  The results are in rough agreement with much more elaborate calculations, both of the purely hadronic variety, and those that invoke quark-gluon degrees of freedom. The quark-gluon and purely hadronic calculations seemingly leading to very similar predictions motivates the question as to whether the two descriptions are dual to each other in the sense of being alternate models, each sufficient to explain the observed data.  There are other indications in the published literature that would seem to motivate the same question.

Kapusta, Lichard, and Siebert \cite{Kapusta:91} calculate the photon emission from a hot hadronic matter and from a quark-gluon plasma at temperatures from $\frac{1}{2}$ GeV to several GeV and find them to be the same.  Scherer and Fearing \cite{Scherer:94} show that different off-shell form factors for s- and u- channel pole diagram contributions to pion compton scattering, calculated in chiral perturbation theory, lead to the same S-matrix, in accordance with Haag's Theorem \cite{Haag:58}.  

Haag proved that in a theory with an S-Matrix (which excludes both QED and QCD) there is no unique off-mass-shell extension.  The necessary conditions for this lack of uniqueness are unknown, to the best of my knowledge.  It may be that modern high-energy experiments, which count jets, rather than isolated free particles, avoid Haag's theorem and admit the presence of unique Lagrangians. Whether the study of heavy ion collisions can help explore the reach of Haag's theorem remains to be seen. 
\newpage
\appendix
\section{RHO-MASS DETAILS}
Summation over the Matsubara modes and evaluation of the angular integrals in Eq \ref{eq:mass1} then gives, for $\rho$-mesons at rest 
\begin{eqnarray}
\mathcal{M}(\vec{p}=0)^{2} & = & M^{2} - \frac{g^{2}}{12\pi ^{2}}\int_{0}^{\infty }k^{2}dk \nonumber \\
                           &   & \{ \int_{0}^{\beta }dy e^{-ip_{0}y}\frac{k^{2}cosh[\omega (\beta-2y)]+1}{\omega ^{2}sinh(\beta \omega /2)^{2}}+ \nonumber \\
                           &   & \frac{(6-\frac{k^{2}}{M^{2}})coth(\beta E/2)}{2E}+\frac{3coth(\beta \omega /2)}{4\omega } + \frac{kcoth(k\beta /2))}{2M^{2}}\} \nonumber \\
\label{eq:mass2}
\end{eqnarray}
$E$ and $\omega $ in Eq \ref{eq:mass2} are the respective energies of a $\rho $ or $pion$ having spatial momentum of magnitude $k$.  The $y$ integration in the preceding equation results in a term $\frac{2\omega coth(\beta \omega /2)}{\omega ^{2} + p_{0}^{2}/4}$.  The $coth(x)$ functions in the equation contain both zero temperature parts and temperature dependent parts according to the identity $coth(\frac{x}{2})= 1 +2n(x)$, where $n(x)$ is the Bose-Einstein distribution function $(e^{x}-1)^{-1}$.  The real zero-temperature part contributes to the infinite mass renormalization and is incorporated into the constant term $M^{2}$, (see \emph{e.g.} \cite{Kap:89}, p.40).  With the understanding that the constant term refers to the renormalized mass, then (and noting that the last term in Eq. \ref{eq:mass2} can be evaluated in closed form), the expression to be evaluated numerically becomes
\begin{eqnarray}
\mathcal{M}(\vec{p}=0)^{2} & = & M^{2} - \frac{g^{2}}{6\pi ^{2}}\int_{0}^{\infty }k^{2}dk \nonumber \\
                           &   & \{\frac{2k^{2}n(\omega \beta )}{\omega [\omega ^{2}+p_{0}^{2}/4]}+\frac{(6-\frac{k^{2}}{M^{2}})n(\beta E)}{2E}+\frac{3n(\beta \omega )}{4\omega }\} + \frac{\pi ^{4}T^{4}}{15M^{2}} \nonumber \\
\label{eq:mass}
\end{eqnarray}

I obtain the width by continuing the fourth component of the $\rho$ momentum four-vector to a positive imaginary plus a real infinitesimal, contributing $\pi $ times a delta function multiplying the first term of the integrand in Eq.\ref{eq:mass2}.  The result is
\begin{equation}
M\Gamma = Im \mathcal{M}^{2} =\frac{1}{12}\frac{g^{2}}{4\pi }M^{2}(1-4(\frac{m}{M})^{2})^{\frac{3}{2}}[1 + 2n(M\beta )].
\label{eq:width}
\end{equation}
The zero-temperature width given by Eq. \ref{eq:width} agrees with that given by Nishijima \cite{Nish} and determines the value of the gauge coupling constant $g$.  The temperature dependent part of the equation predicts that the $\rho $ width broadens with temperature when the $\rho $ is immersed in a heat bath due to the increased density of final states.
 
\section{EQUATION OF STATE CALCULATION}
The $\rho $-propagator $D(p)^{\mu \nu }$ in Eq \ref{eq:A} has two poles on the real $p_{0}$-axis, at $|\vec{p}|$ and at $E(p)$.  It is convenient to separate the poles, so that the propagator $D(p)^{\mu \nu }$ becomes $\frac{1}{M^{2}} [\frac{1}{p^{2}}-\frac{1}{p^{2}+M^{2}}]p^{2}L(p)^{\mu \nu }$. After doing the Matsubara sums and the the angular integrals the resulting equation for the second order (in $g$) contribution to the dimensionless pressure, $\frac{P^{2}}{T^{4}}$, becomes
\begin{equation}                                                                            
P^{(2)}/T^{4}=-\beta ^{3}\frac{6}{\pi ^3}(\frac{g^{2}}{4\pi })\int_{0}^{\infty }kdk\{ \frac{n[\beta \omega (k)]}{\omega (k)}X(T,k)+\frac{n[\beta E(k)]}{E(k)}Y(T.k) \}
\label{eq:pt4}
\end{equation}
where
\begin{eqnarray}
X(T,k) & = & \frac{-\pi ^{4}}{15}T^{4}\frac{k^{2}}{M^{2}}+ \int_{0}^{\infty }qdq\{ \nonumber \\
       &   & \\frac{n[\beta \omega (q)]}{\omega (q)}[\frac{(M^{2}-4\mu ^{2})}{8}\ln R -kq]+	\nonumber \\
       &   & \frac{n[\beta E(q)]}{E(q)}[\frac{[(M^{2}-4\mu ^{2})}{8}\ln S +kq(1+\frac{q^{2}}{M^{2}}]  \nonumber \\
\label{eq:question}
\end{eqnarray}
and
\begin{equation}
Y(T,k)=\frac{-3}{2}\frac{kn[\beta E(k)]}{E(k)}\int_{0}^{\infty}q^(2)dq\{\frac{n[\beta \omega (q)]}{\omega (q)} + 2\frac{n[\beta E(q)]}{E(q)}\}
\label{eq:expl}
\end{equation}
The two arguments of the log functions are $R=\frac{(M^{2}-4\mu ^{2})E(k+q)^{2}+(k-q)^{2}}{(M^{2}-4\mu ^{2})E(k-q)^{2}+(k+q)^{2}}$
 and $S=\frac{(M^{2}-4\mu ^{2})E(q)^{2}+(q+2k)^{2}}{(M^{2}-4\mu ^{2})E(q)^{2}+(q-2k)^{2}}$. 
.  
\section{ACKNOWLEDGEMENTS}
I am indebted to the Argonne National LaboratoryHigh Energy Physics Lunch Seminar, where the substance of this paper was first presented on April 19, 2005, and to the following for helpful comments: Geoffrey Bodwin, Ralf Rapp, Hal Spinka, Tim Tait, Arthur Weldon, and also to Cosmas Zachos for a critical reading of an earlier draft and for first suggesting the concept of duality in the context of this investigation.  I am grateful to Proffessors Rapp and Wambach for permission to copy their P-T graph from ref. \cite{Rapp:95}.  Work in the High Energy Physics Division at Argonne is supported by The U. S. Department of Energy, Division of High Energy Physics, Contract No. W-31-109-ENG-38.

\end{document}